\documentclass[A4paper,superscriptaddress,twocolumn,aps,floatfix,citeautoscript,nobibnotes]{revtex4-1} 
\usepackage{epsfig,epstopdf}
\usepackage{graphicx}
\usepackage{amsfonts,amsmath,amssymb,mathtools}
\usepackage{appendix}
\usepackage{latexsym}
\usepackage{bm}
\usepackage{cases}
\usepackage{color}
\usepackage[colorlinks=true,allcolors=blue,breaklinks=true]{hyperref}
\usepackage{multirow}
\usepackage{soul}
\usepackage[utf8]{inputenc}

\newcommand{\zeq}{\zeta_{\mathrm{eq}}}
\newcommand{\zdep}{\zeta_{\mathrm{dep}}}
\newcommand{\zth}{\zeta_{\mathrm{th}}}

\newcommand{\lopt}{\ell_{\mathrm{opt}}}
\newcommand{\lav}{\ell_{\mathrm{av}}}
\newcommand{\qav}{q_{\mathrm{av}}}
\newcommand{\qopt}{q_{\mathrm{opt}}}
\newcommand{\nueq}{\nu_{\mathrm{eq}}}
\newcommand*\un[1]{\,\mathrm{#1}}

\begin{document}

\title{Field-dependent roughness of moving domain walls in a Pt/Co/Pt magnetic thin film}

\author{María José Cortés Burgos}
\affiliation{Instituto de Nanociencia y Nanotecnología, CNEA-CONICET, Centro Atómico Bariloche, Av. E. Bustillo 9500, R8402AGP San Carlos de Bariloche, Río Negro, Argentina}
\affiliation{Instituto Balseiro, Universidad Nacional de Cuyo - CNEA, Av. E. Bustillo 9500, R8402AGP San Carlos de Bariloche, Río Negro, Argentina}

\author{Pamela C. Guruciaga}
\altaffiliation[Currently at ]{European Molecular Biology Laboratory (EMBL), 69117 Heidelberg, Germany.}
\affiliation{Centro Atómico Bariloche, Comisión Nacional de Energía Atómica (CNEA), Consejo Nacional de Investigaciones Científicas y Técnicas (CONICET), Av. E. Bustillo 9500, R8402AGP San Carlos de Bariloche, Río Negro, Argentina}

\author{Daniel Jordán}
\affiliation{Instituto Balseiro, Universidad Nacional de Cuyo - CNEA, Av. E. Bustillo 9500, R8402AGP San Carlos de Bariloche, Río Negro, Argentina}

\author{Cynthia P. Quinteros}
\altaffiliation[Currently at ]{Escuela de Ciencia y Tecnología, Universidad de San Martín, B1650KNA Buenos Aires, Argentina. CONICET.}
\affiliation{Instituto de Nanociencia y Nanotecnología, CNEA-CONICET, Centro Atómico Bariloche, Av. E. Bustillo 9500, R8402AGP San Carlos de Bariloche, Río Negro, Argentina}

\author{Elisabeth Agoritsas}
\affiliation{Institute of Physics, Ecole Polytechnique Fédérale de Lausanne (EPFL), CH-1015 Lausanne, Switzerland}

\author{Javier Curiale}
\affiliation{Instituto de Nanociencia y Nanotecnología, CNEA-CONICET, Centro Atómico Bariloche, Av. E. Bustillo 9500, R8402AGP San Carlos de Bariloche, Río Negro, Argentina}
\affiliation{Instituto Balseiro, Universidad Nacional de Cuyo - CNEA, Av. E. Bustillo 9500, R8402AGP San Carlos de Bariloche, Río Negro, Argentina}

\author{Mara Granada}
\affiliation{Instituto de Nanociencia y Nanotecnología, CNEA-CONICET, Centro Atómico Bariloche, Av. E. Bustillo 9500, R8402AGP San Carlos de Bariloche, Río Negro, Argentina}
\affiliation{Instituto Balseiro, Universidad Nacional de Cuyo - CNEA, Av. E. Bustillo 9500, R8402AGP San Carlos de Bariloche, Río Negro, Argentina}

\author{Sebastian Bustingorry}
\affiliation{Instituto de Nanociencia y Nanotecnología, CNEA-CONICET, Centro Atómico Bariloche, Av. E. Bustillo 9500, R8402AGP San Carlos de Bariloche, Río Negro, Argentina}

\date{\today}

\begin{abstract}
    The creep motion of domain walls driven by external fields in magnetic thin films is described by universal features related to the underlying depinning transition. One key parameter in this description is the roughness exponent characterizing the growth of fluctuations of the domain wall position with its longitudinal length scale. The roughness amplitude, which gives information about the scale of fluctuations, however, has received less attention. Albeit their relevance, experimental reports of the roughness parameters, both exponent and amplitude, are scarce. We report here experimental values of the roughness parameters for different magnetic field intensities in the creep regime at room temperature for a Pt/Co/Pt thin film. The mean value of the roughness exponent is $\zeta = 0.74$, and we show that it can be rationalized as an effective value in terms of the known universal values corresponding to the depinning and thermal cases. In addition, it is shown that the roughness amplitude presents a significant increase with decreasing field. These results contribute to the description of domain wall motion in disordered magnetic thin systems. 
\end{abstract}

\maketitle

\section{Introduction}

Domain wall motion in thin magnetic films has long been studied. It comprises the action of external drives as magnetic fields or electrical currents; it also involves thermal effects and a key role of disorder. Since the seminal work of Lemerle \textit{et al.}~\cite{lemerle_domainwall_creep} relating domain wall motion to universal characteristics of disordered systems, there have been a lot of studies tackling this problem both from theoretical~\cite{chauve_creep_long, chauve_creep_short, agoritsas_review, martinez2008, Martinez2012, kolton_creep2, kolton_dep_zeroT_long, kolton_depinning_zerot2, kolton_short_time_exponents, rosso_dep_exponent, rosso_depinning_longrange_elasticity, duemmer2, bustingorry_thermal_rounding_epl, Bustingorry2009, bustingorry_thermal_rouding_long, bustingorry_thermal_depinning_exponent, Ferrero2013nonsteady, Ferrero2013, caballero2017excess, caballero2018magnetic} and experimental~\cite{bauer_deroughening_magnetic2, metaxas_depinning_thermal_rounding, Moon2013, Lee2011, Ryu2011, Gorchon2014, Jeudy2016, jeudy2018pinning, ferre2013universal, wells2017, savero2019universal, quinteros2018, shahbazi2018magnetic, pardo2017} perspectives.

Intrinsic disorder induces a zero temperature depinning transition~\cite{fisher_depinning_meanfield}. Consider a magnetic thin film with perpendicular anisotropy such that domains with the magnetization pointing in opposite out-of-plane directions are separated by a domain wall. The motion of the domain wall can be induced by an external out-of-plane magnetic field $H$ favoring one of the magnetization domains. Ideally, at zero temperature, the domain wall does not move when the external field is small because it is pinned by the disorder. A critical field value, the depinning field $H_d$, must be reached in order to force the domain wall to acquire a finite velocity $v$. For $H \gg H_d$ velocity increases linearly with the field, in the so-called fast-flow regime. 
Approaching the critical field from above, the velocity vanishes following a power-law behavior $v \sim (H-H_d)^\beta$, with $\beta$ the velocity critical exponent.
Finite temperature values break down the abrupt transition, allowing for finite velocities even below $H_d$. On one hand, a finite temperature value rounds the abrupt depinning transition exactly at $H_d$, with the velocity vanishing as $v \sim T^\psi$, with $\psi$ the thermal rounding critical exponent~\cite{bustingorry_thermal_rounding_epl}. On the other hand, thermal activation well below $H_d$ gives rise to creep motion characterized by a non linear increase of the velocity as $v \sim \exp(-H^{-\mu})$, with $\mu$ the universal creep exponent~\cite{nattermann_creep_law, chauve_creep_long}. This is the complex velocity-field response scenario for domain wall motion in a disordered magnetic thin film.

The information carried by the set of universal critical exponents ($\beta$, $\psi$ and $\mu$) associated to dynamic aspects of the depinning transition is complemented by the set of roughness exponents characterizing fluctuations in the position of the domain wall. In general, the roughness exponent $\zeta$ characterizes the growth of transverse fluctuations of the domain-wall position as a function of the longitudinal length scale $r$. These fluctuations can be defined in terms of a displacement-displacement correlation function as the roughness function $B(r)$ or the structure factor $S(q)$, both defined below, such that $B(r) \sim r^{2 \zeta}$ or ${S(q) \sim q^{-(1+2\zeta)}}$. In terms of such geometrical fluctuations, three reference states can be defined associated to the equilibrium ($H = 0$), depinning ($H = H_d$) and fast-flow ($H \gg H_d$) states, each of them characterized by different values of the roughness exponent: $\zeq$, $\zdep$ and $\zth$, respectively~\cite{kolton_dep_zeroT_long}.
In principle, at any finite field value $H$, fluctuations of the domain-wall position at a given length scale $r$ can be described in terms of these different roughness exponents.

The work of Lemerle \textit{et al.}~~\cite{lemerle_domainwall_creep} reported the first experimental value for the roughness exponent associated to the creep regime of domain wall motion. They reported $\zeta = 0.69 \pm 0.07$, which was at first interpreted as the equilibrium roughness exponent $\zeq = 2/3$. However, this is not compatible with more recent predictions~\cite{kolton_dep_zeroT_long} since the scale of the experiment is well above the length scale where the equilibrium roughness exponent should be observed. 
Following the work by Lemerle \textit{et al.}~\cite{lemerle_domainwall_creep}, experimental reports of roughness exponents show a wide range of reported values~\cite{Shibauchi2001, Huth2002, Lee2009, Moon2013, savero2019universal, DiazPardo2019, domenichini2019transient, Jordan2020, Albornoz2021}. For field-induced motion of one-dimensional domain walls, reported values can be found in the range $0.60 < \zeta < 0.83$ for Pt/Co/Pt, around $\zeta = 0.6$ for (Ga,Mn)(As,P), and {$0.72 < \zeta < 0.82$ in GdFeCo~\cite{Albornoz2021}}. Although the values of $\zeta$ were initially interpreted as equilibrium values, how to rationalize the wide range of observed values remains as an open question.

In addition to these exponent values, the amplitude of the roughness function $B(r)$ is a direct measure of the typical scale of the fluctuations and has received much less attention, most likely because it is not a universal quantity. The roughness amplitude, in principle, depends on the temperature of the system and on the elasticity of the domain wall~\cite{agoritsas2010, agoritsas_review}. However, recent experimental observations of the roughness amplitude indicate that it depends also on the applied magnetic field~\cite{Jordan2020, Albornoz2021}. Therefore, a more detailed study of the dependence of the roughness amplitude on external parameters is indeed needed.

In a recent study, the interpretation of roughness parameters in terms of length scales crossovers has been presented~\cite{Albornoz2021}. Ferrimagnetic GdFeCo samples were studied focusing on different temperature and field conditions, but a systematic field-dependent study is lacking.
Here, we report on the field dependence of the roughness parameters, both the roughness exponent and the roughness amplitude, well in the creep regime of motion of domain walls in a Pt/Co/Pt magnetic thin film. We pay special attention to the experimental protocol so as to perform all the measurements in exactly the same conditions. We obtain that the roughness exponent is almost constant when varying the magnetic field, with a mean value $\zeta = 0.74$. We explain this result as an effective value which results from a finite crossover length scale between depinning roughness and thermal roughness reference states -- characterized by $\zdep$ and $\zth$ respectively. In addition, the roughness amplitude is shown to increase with decreasing magnetic field, which suggests that transverse fluctuations at the Larkin scale (above which disorder dominates over elasticity) are field-dependent.
The rest of the manuscript is organized as follows. Section~\ref{sec:exp-details} presents experimental details about the sample, the image acquisition of domain walls and a preliminary characterization of the velocity-field behavior. Then, the main experimental results and their analysis are presented in Sec.~\ref{sec:exp-results}. Section~\ref{sec:analysis} presents an interpretation of the experimental data in terms of a crossover between roughness regimes at different scales. The findings are then analyzed using scaling relations in Sec.~\ref{sec:scaling}, with details deferred to the Appendix~\ref{sec:app}. Finally, Secs.~\ref{sec:disc} and \ref{sec:conc} are devoted to a discussion of the results and the conclusion of the work, respectively.

\section{Experimental details}
\label{sec:exp-details}

We studied the roughness of domain walls in a Pt(8~nm)/Co(0.4~nm)/Pt(4~nm) magnetic thin film with perpendicular magnetization. The sample was deposited on a silicon substrate by dc magnetron sputtering at room temperature in a $(2.8 \pm 0.1)\times 10^{-3}\un{Torr}$ Ar atmosphere (see Ref.~\cite{quinteros2020} for further details).

A home-made polar magneto optical Kerr effect (PMOKE) microscope was used to image magnetic domains. Relevant components of the microscope are a $5\times$ objective (\textit{Olympus} LMPLFLN series), a 637 nm dominant wavelength high-brightness LED, two Glan-Thompson polarizers, and an EXi Blue ($1392\times1040$ pixels, 14 bit) CCD from QImaging Corp. The illumination of the PMOKE microscope was set in a Köhler configuration and we used a $2\times2$ binning that gives an effective pixel size of $\delta = 0.936\un{\mu m}$.

\begin{figure}[tb]
    \includegraphics[width=\linewidth]{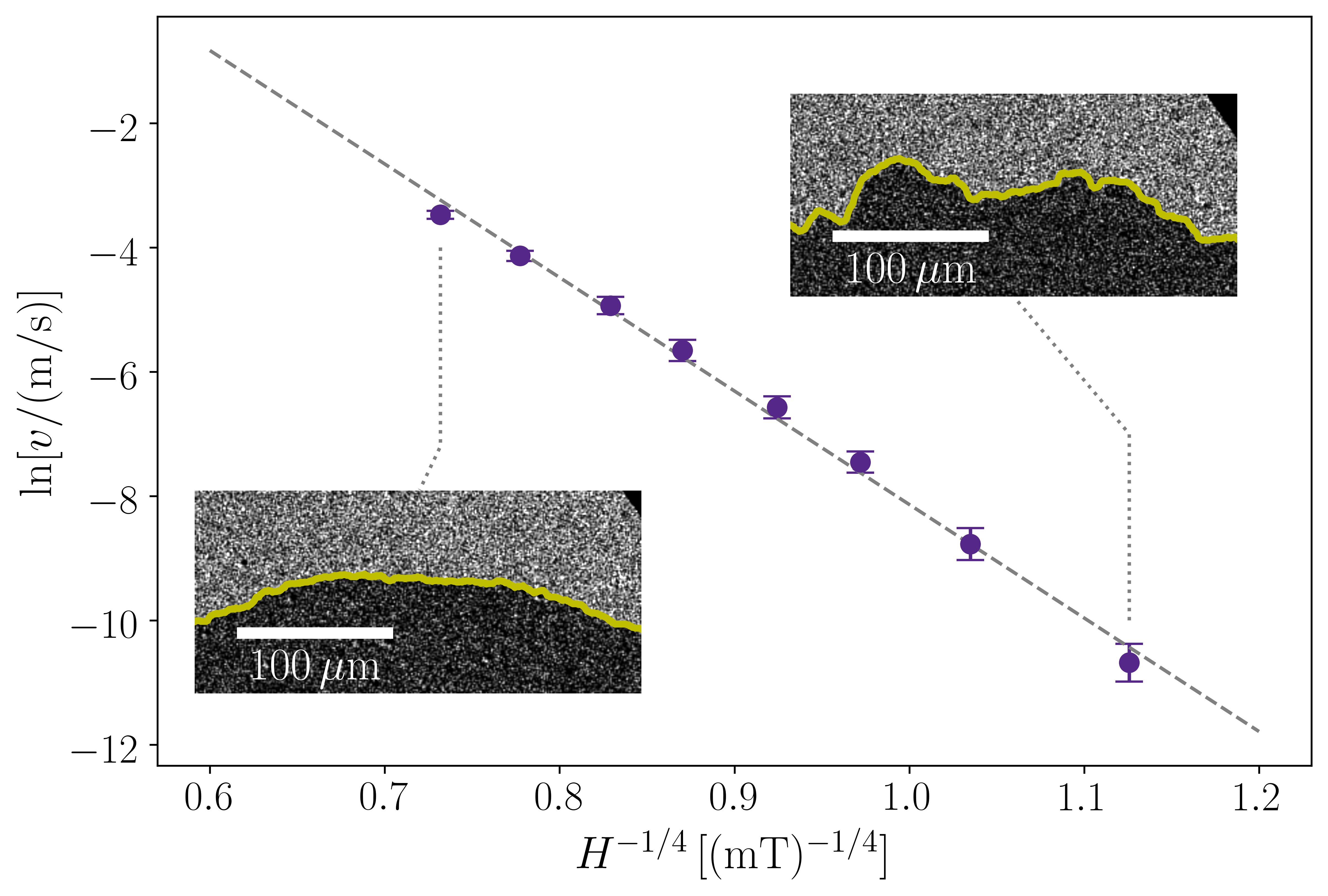}
\caption{Creep plot of the velocity,  $\ln v$ vs. $H^{-1/4}$. The linear behavior indicates that, for the studied field range, the system is in the creep regime. Two domain wall profiles are shown, corresponding to the maximum and minimum velocities.}
    \label{fig:creep}
\end{figure}

The magnetization was first saturated in an out-of plane direction and then a magnetic domain was nucleated by applying a magnetic field pulse in the opposite direction.
Once a domain wall was nucleated, a magnetic field pulse of intensity $H$ and duration $\Delta t$ was applied to move the domain wall. Differential images were used to measure domain-wall velocities, which were computed as $v = \Delta x/ \Delta t $, where $\Delta x$ was the average advance of the domain wall. We used magnetic fields in the range $0.6\un{mT} < H < 3.5\un{mT}$ and pulse durations $0.001\un{s} < \Delta t < 1\un{s} $ to measure velocities well in the creep regime. Figure~\ref{fig:creep} shows the velocity against field in a creep plot, $\ln (v)$ vs. $H^{-1/4}$. The linear behavior observed for velocities in the range $ 23\un{\mu m/s} < v < 31\un{mm/s}$ indicates a good agreement with the creep regime~\cite{Jeudy2016},
\begin{equation}
\label{eq:creep-law}
    \ln v = \ln v_d + \frac{T_d}{T} - \frac{T_d H_d^{1/4}}{T} H^{-1/4},
\end{equation}
where $v_d$, $T_d$ and $H_d$ are the depinning velocity, depinning temperature and depinning field, respectively (see Ref.~\cite{jeudy2018pinning} and references therein).
Using the linear fit (dashed line in Fig.~\ref{fig:creep}) and considering the depinning velocity $v_d$ for Pt/Co/Pt thin films to be in the range $1\un{m/s}-20\un{m/s}$~\cite{jeudy2018pinning}, the depinning temperature and depinning field can be estimated using Eq.~\eqref{eq:creep-law} as $T_d = (2400 \pm 400)\un{K}$ and $H_d = (25 \pm 15)\un{mT}$, respectively. These rough estimates will serve to assess important length scales below.

The characterization of domain wall roughness was performed with fixed and controlled experimental conditions. The range of explored velocities was set by the use of a single coil to produce the magnetic field. Larger velocities could be produced by a different (smaller) coil which we did not use to avoid possible changes in the homogeneity of the field. The temperature of the sample was stabilized at 295 K.
Furthermore, the same physical region of the sample was used to acquire images, with the obtained domain wall profiles corresponding to different positions in the field of view.
The two insets in Fig.~\ref{fig:creep} show domain wall profiles obtained after applying magnetic field pulses of different intensities, corresponding to the maximum and minimum velocities. The different gray levels correspond to out-of-plane magnetization in opposite directions. 

In order to evaluate the roughness of domain walls, it is necessary to obtain good statistics for each set of parameters~\cite{Jordan2020}.
A sequence of magnetic field pulses with the same intensity and pulse duration was used, assuring the domain wall stays within the field of view. A set of $N$ images, all under the same experimental conditions, were taken after each pulse (as indicated in Fig.~\ref{fig:profiles}, where the average position was subtracted to every domain wall profile).
In addition, all imaged domain walls were cropped using the same longitudinal size $L = M \delta$ with $M = 330$, resulting in $L \approx 309\un{\mu m}$. Since finite size effects are important when computing the roughness~\cite{Jordan2020}, having the same size $L$ for all profiles allows us to average properly.

\section{Experimental roughness parameters}
\label{sec:exp-results}

\begin{figure}[tb]
    \includegraphics[width=\linewidth]{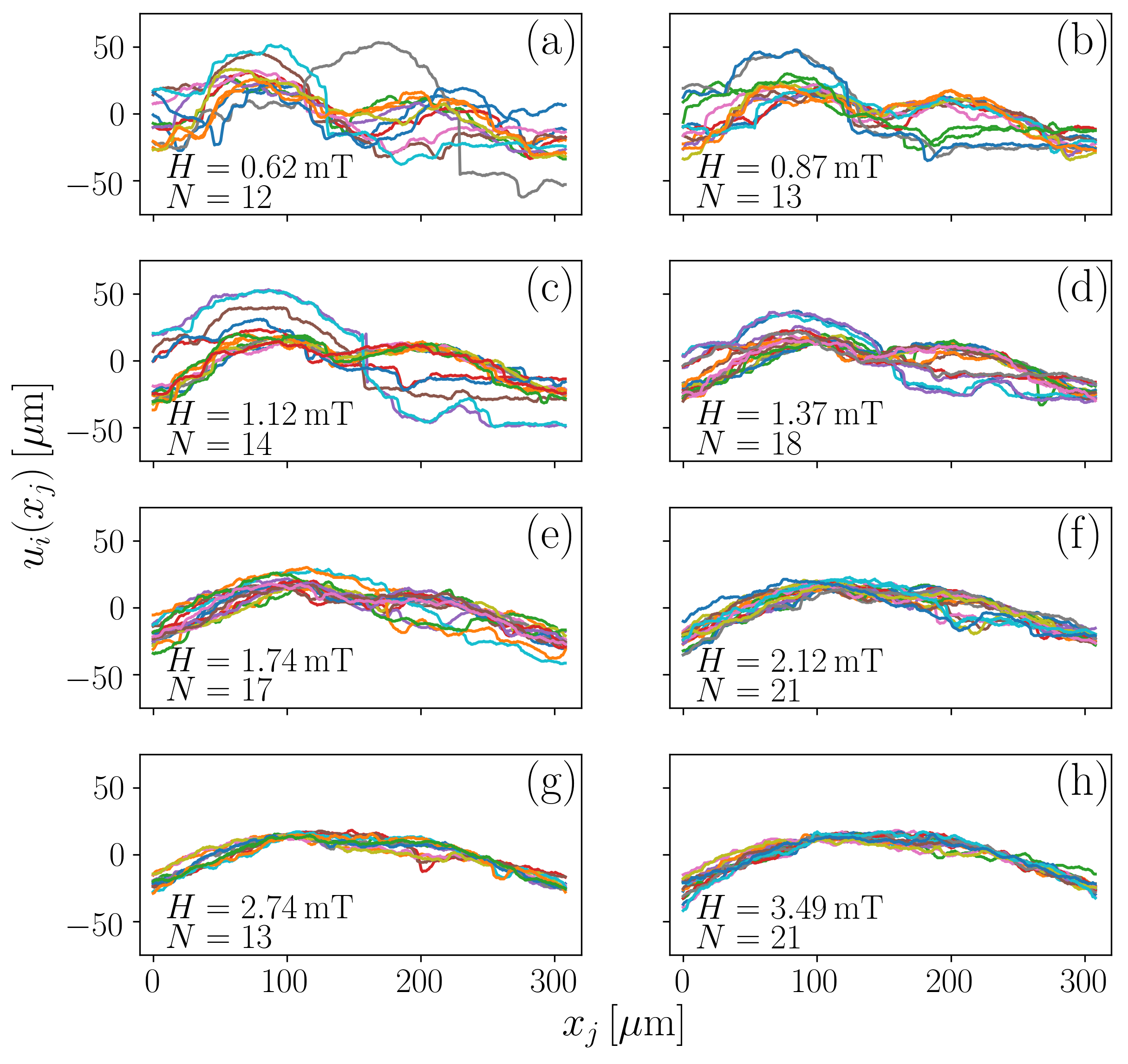}
    \caption{Domain wall profiles used to evaluate the roughness properties. For each field the longitudinal length $L$ is fixed but the number of individual profiles $N$ varies, as indicated in each panel (a-h).}
    \label{fig:profiles}
\end{figure}

From PMOKE images we obtained domain wall positions, which we represent as $u_i(x_j)$, with $x_j = \delta j$ and $j=1,...,M$. The index $i$ runs over the $N$ profiles obtained for each field value. All profiles used in this work, with the same longitudinal length $L$, are shown in Fig.~\ref{fig:profiles}. The number of profiles for each field, $N$, is indicated in the figure. These profiles were taken in different positions within the same field of view of the PMOKE microscope, thus $x_j$ is referred to the initial point of the domain wall and the variable $u_i(x_j)$ is shifted such that its mean value is zero.

\begin{figure}[tb]
    \includegraphics[width=\linewidth]{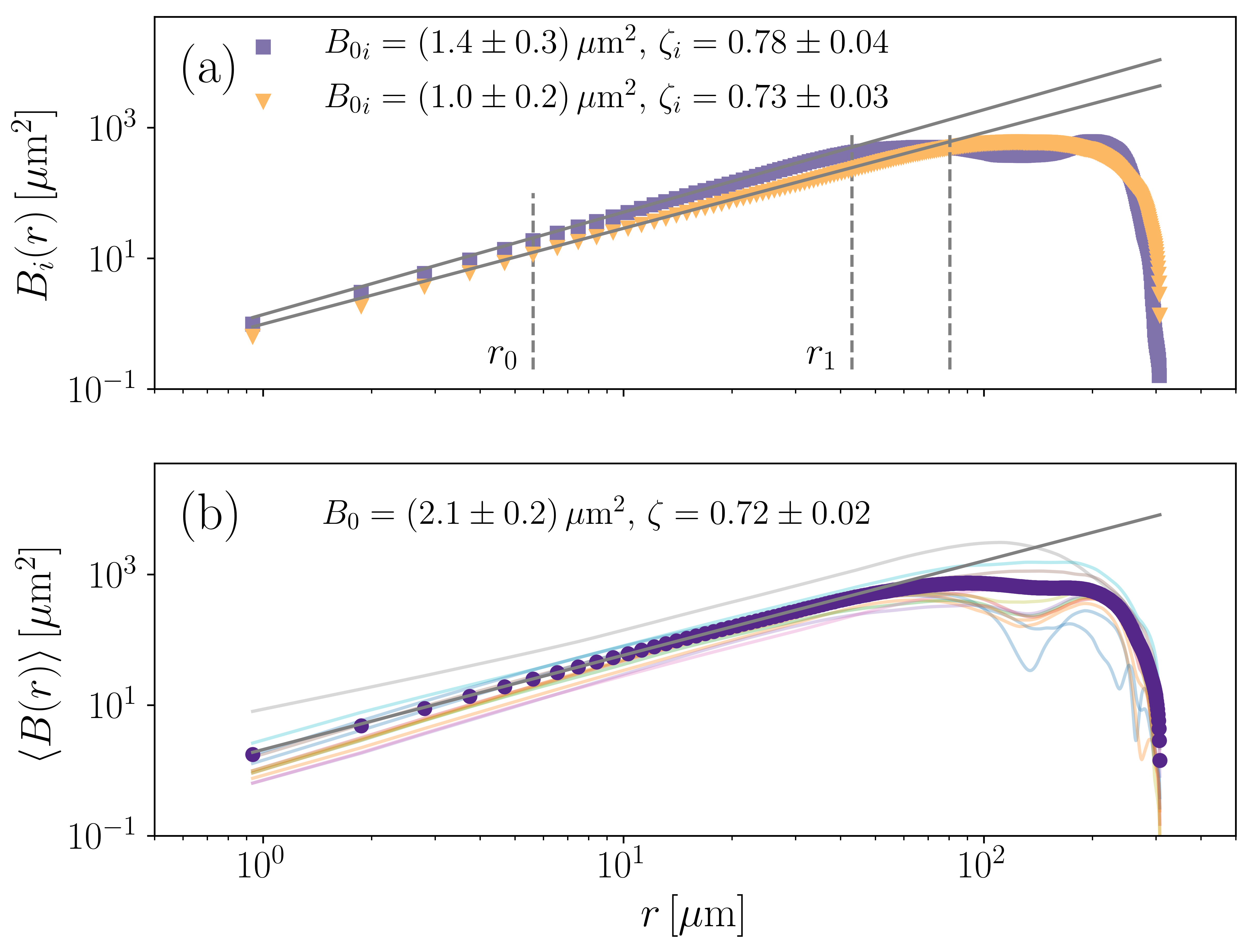}
    \caption{(a) Individual roughness functions $B_i(r)$ for two different profiles obtained using $H = 0.62\un{mT}$. The individual roughness parameters $\zeta_i$ and ${B_0}_i$ for each case are indicated. $r_0$ and $r_1$ correspond to the boundaries obtained from the fitting procedure.
    (b) Average roughness function $\langle B(r) \rangle$ corresponding to $H = 0.62\un{mT}$. The roughness parameters $\zeta$ and $B_0$ are indicated. Thin light continuous lines correspond to the $N = 12$ different individual roughness functions $B_i(r)$, obtained using the profiles in Fig.~\ref{fig:profiles}(a), used to obtain the average value.}
    \label{fig:Bder}
\end{figure}

\begin{figure}[tb]
    \includegraphics[width=\linewidth]{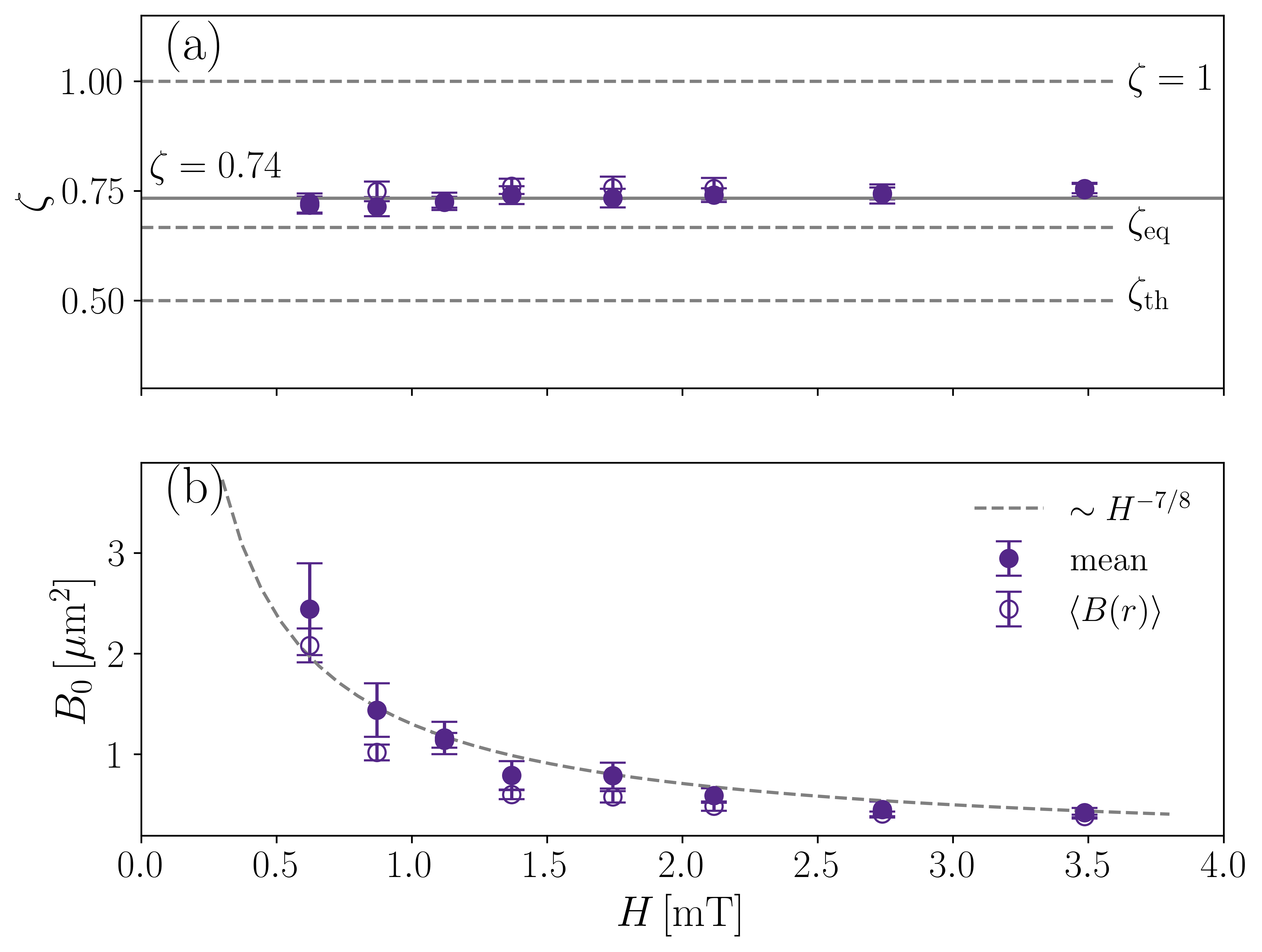}
    \caption{Field dependence of the roughness parameters. 
    (a) Roughness exponent $\zeta$. The three reference values, $\zth$, $\zeq$ and $\zeta = 1$, are indicated as horizontal dashed lines. The continuous horizontal line corresponds to $\zeta = 0.74$, the mean of the obtained experimental values.
    (b) Roughness amplitude $B_0$.
    Reported values obtained as the mean of the individual roughness parameters and as the roughness parameters of the average roughness function $\langle B(r) \rangle$, as indicated in the key. The dashed line indicated the $B_0 \sim H^{-\theta}$ behavior, with $\theta = 7/8$.}
    \label{fig:roughness-parameters}
\end{figure}

Fluctuations of domain-wall position can be accounted for by the roughness function, a displacement-displacement correlation function. For a given profile $u(x_j)$ the roughness function $B(r)$ is defined as
\begin{equation}
\label{eq:Br}
    B(r) = \frac{1}{M-k}\sum_{j=1}^{M-k} \left[ u(x_j+r) - u(x_j)\right]^2,
\end{equation}
where $k = r/\delta < M$ is an integer value.
Notice that, by construction, the statistics of $B(r)$ decreases for large $r$ values. However, good statistics is obtained for small $r$, which is in the range where the roughness parameters are extracted (see below).
For self-affine profiles a power-law growth of $B(r)$ is expected,
\begin{equation}
\label{eq:fitBr}
    B(r)=B_0\left(\frac{r}{\ell_0}\right)^{2\zeta} \,,
\end{equation}
where $\zeta$ is the roughness exponent and $B_0$ is the roughness amplitude. We measure $r$ in $\un{\mu m}$ and set $\ell_0 = 1\un{\mu m}$, so that $B_0$ has the same units as $B(r)$. Therefore, using the fitting procedure described in Ref.~\cite{Jordan2020} we can extract the roughness parameters for each individual domain wall profile, obtaining both $\zeta_i$ and ${B_0}_i$.
In order to illustrate how the roughness parameters of individual domain-wall profiles fluctuate, Fig.~\ref{fig:Bder}(a) shows two $B_i(r)$ functions corresponding to two different profiles obtained with the same field $H = 0.62\un{mT}$. The roughness parameters and the boundaries of the fitting range obtained as part of the fitting procedure for each case are indicated.
The boundaries $r_0$ and $r_1$ are obtained as those guaranteeing that the goodness of the fit, measured using the coefficient of determination, is above $R^2 = 0.9995$ when fitting using a finite range inside $(r_0,r_1)$. The same value for the lower bound of $R^2$ was used to analyze all experimental profiles.
The reported values for the roughness parameters represent the best values taking into account possible variations in the limits of a linear fit inside $(r_0,r_1)$~\cite{Jordan2020}.
Finite-size effects, evidenced by the drop of the roughness function at large scales, are taken into account in the value of $r_1$.
From the individual values of roughness parameters $\zeta_i$ and ${B_0}_i$, the mean values $\langle \zeta \rangle$ and $\langle B_0 \rangle$ can be obtained for each set of profiles in Fig.~\ref{fig:profiles}. These are shown in Fig.~\ref{fig:roughness-parameters} (solid dots) as functions of the applied field.

In addition, the average roughness function $\langle B(r) \rangle$ can be computed from the roughness functions $B_i(r)$ obtained from the individual roughness profiles. This is possible because the same system size $L$ is used for all profiles. Otherwise, $B_i(r)$ for large $r$ values would not have the same statistical weight and finite-size effects would not be under control. The average roughness function is also expected to recreate self-affine behavior, with $\langle B(r) \rangle \sim r^{2 \zeta}$. We show in Fig.~\ref{fig:Bder}(b) the average roughness function corresponding to $H = 0.62\un{mT}$, together with all the $N = 12$ individual $B_i(r)$ roughness functions. A linear fit of $\langle B(r) \rangle$ was performed using the same protocol as for the individual $B_i(r)$. The obtained values for the roughness parameters, $\zeta$ and $B_0$, are shown in Figs.~\ref{fig:roughness-parameters} (empty dots).

We have then two sets of roughness parameters, obtained as the mean values of individual fitting parameters, $\langle \zeta \rangle$ and $\langle B_0 \rangle$, or as the fitting parameters of the average roughness functions $\langle B(r) \rangle$, $\zeta$ and $B_0$. These values are expected to converge to the same value for large $N$. Figure~\ref{fig:roughness-parameters} shows the field dependence of the two sets of roughness parameters, where a good agreement is observed between the two sets of values.
As shown in Fig.~\ref{fig:roughness-parameters}(a) the roughness exponent does not change noticeably with the field. The mean value $\zeta = 0.74$ is indicated, which is obtained using all values in Fig.~\ref{fig:roughness-parameters}(a).
Quite differently, the roughness amplitude $B_0$ is decreasing with increasing field $H$, as observed in Fig.~\ref{fig:roughness-parameters}(b). This quantitative change accounts for the qualitative change in the roughness observed in the profiles shown in Fig.~\ref{fig:profiles}, as discussed in Ref.~\cite{Jordan2020}.

We also include in Fig.~\ref{fig:roughness-parameters}(a) the exponent values expected for the reference states in the quenched Edwards-Wilkinson universality class: $\zeq = 2/3$, $\zeta = 1$ and $\zth = 1/2$, corresponding to the equilibrium, depinning and thermal (or fast-flow) regimes, respectively~\cite{kolton_dep_zeroT_long}.
Strictly speaking, in the depinning reference case the structure factor is $S(q)\sim q^{-(1+\zdep)}$, with $\zdep=1.25$; but in such a super-rough case (i.e. an exponent larger than $1$) the roughness function $B(r)$ exhibits an effective exponent $\zeta=1$~\cite{Lopez1997}, hence the value shown in Fig.~\ref{fig:roughness-parameters}(a). The discrepancy between these reference states and the measured exponent is the main point discussed in the next section.

\section{Crossovers between roughness regimes}
\label{sec:analysis}

The values for the roughness exponents reported in Fig.~\ref{fig:roughness-parameters}(a) do not comply with the expected reference values for a domain wall moving in a disordered medium, within the quenched Edwards-Wilkinson universality class (short-range elasticity, random-bond quenched disorder, and external field~\cite{Ferrero2013}). As shown in Fig.~\ref{fig:roughness-parameters}(a) the mean value for the roughness exponent is $\zeta = 0.74$, which is different from $\zth$, $\zeq$ and $\zeta = 1$. This situation is common to other experimental results~\cite{Shibauchi2001, Huth2002, Lee2009, Moon2013, savero2019universal, DiazPardo2019, domenichini2019transient, Jordan2020, Albornoz2021}. In order to reconcile the experimentally obtained value for the roughness exponent with the theoretical expected values, the crossover length scales between different roughness regimes should be considered~\cite{kolton_dep_zeroT_long, Albornoz2021}.

Domain-wall position fluctuations can also be described in Fourier space using the structure factor, which is usually preferred in theoretical developments. Given a generic domain-wall profile $u(x_j)$, the structure factor can be written as
\begin{equation}
    S(q_n) = \left| \frac{1}{M} \sum_{j=1}^{M} u(x_j) e^{-i q_n x_j} \right|^2,
\end{equation}
where $x_j = j \delta$, the system size is $L = M \delta$, and $q_n = 2 \pi n/L$ for $n=1,\ldots,M$.
In the following, keeping in mind that $q_n$ takes discrete values, we drop the sub-index $n$ for simplicity, and refer just to $S(q)$.
If $u(x_j)$ represents a generic profile containing the key roughness features, then $S(q)$ should contain the three roughness regimes at different scales~\cite{kolton_dep_zeroT_long}.
At lengthscales above the Larkin length $L_c$ (i.e. $q<q_c \sim 1/L_c)$, disorder dominates over elasticity and a self-affine behavior of the roughness function is expected.
For a finite magnetic field, we expect the three roughness regimes to be separated by two crossover length scales: $\lopt$, corresponding to the typical size of a domain-wall segment that should be moved in order to overcome the optimal energy barrier in the creep regime~\cite{nattermann_creep_law, chauve_creep_long, kolton_dep_zeroT_long, Jeudy2016}, and $\lav$ the typical size of depinning avalanches~\cite{kolton_dep_zeroT_long}. For length scales $r < \lopt$ equilibrium roughness fluctuations, given by $\zeq$, are expected. In the range $\lopt < r < \lav$ we expect depinning fluctuations, described by $\zdep$. Finally, for $r > \lav$ roughness is given by $\zth$, characteristic of thermal fluctuations in the fast-flow regime.
This can be expressed in terms of the structure factor in a useful way: we expect $S(q) \sim q^{-(1+2\zeq)}$ for $q > \qopt$, $S(q) \sim q^{-(1+2\zdep)}$ for $\qopt > q > \qav$, and $S(q) \sim q^{-(1+2\zth)}$ for $\qav > q$~\cite{kolton_dep_zeroT_long}. The two length scales, $\lopt \equiv 2\pi/\qopt$ and $ \lav \equiv 2\pi/\qav$, should be taken into account to rationalize the experimentally observed value of the roughness exponent~\cite{Albornoz2021}.

In the present case, the length scale associated to the optimal energy barrier, $\lopt$, is below the minimum experimental length scale, $\delta$. Using $L_c = 50\un{nm}$~\cite{jeudy2018pinning} as an estimate for the Larkin length, and the estimated value for the depinning field $H_d = (25 \pm 15)\un{mT}$, the field dependence of $\lopt$ can be obtained using~\cite{chauve_creep_long, kolton_dep_zeroT_long}
\begin{equation}
\label{eq:lopt}
    \lopt = L_c \left( \frac{H}{H_d} \right)^{-\nueq},
\end{equation}
where $\nueq$ is a critical exponent characterizing the divergence of $\lopt$, and $\nueq = 1/(2-\zeq) = 3/4$ in our case~\cite{kardar_review_lines}.
For the field range of the experiment we obtain $ 0.2 \un{\mu m} < \lopt < 0.8 \un{\mu m}$, and then $\lopt(H) \lesssim \delta = 0.936\un{\mu m}$. We consider then that $\lopt$ is smaller than the typical length scales in the experiment and thus equilibrium fluctuations given by $\zeq$ can safely be neglected.
We shall then describe roughness properties in terms of the depinning and thermal reference states, characterized by $\zdep$ at small length scales and $\zth$ at large length scales, respectively.

The structure factor containing a crossover between depinning and thermal fluctuations can be defined as
\begin{equation}
\label{eq:crossfn}
    S(q) = \frac{1}{\dfrac{1}{S_{\mathrm{dep}}(q)} + \dfrac{1}{S_{\mathrm{th}}(q)}},
\end{equation}
with
\begin{eqnarray}
    S_{\mathrm{dep}}(q) &=& \widetilde{S} \left( \frac{q}{\qav} \right)^{-(1+2\zdep)}, \label{eq:Sdep}\\
    S_{\mathrm{th}}(q) &=& \widetilde{S} \left( \frac{q}{\qav} \right)^{-(1+2\zth)}, \label{eq:Sth}
\end{eqnarray}
where $\widetilde{S}$ is a characteristic amplitude, and $\qav = 2 \pi /\lav$ is the crossover wave vector. The  structure factor then has two limiting cases, $S(q) = S_{\mathrm{dep}}(q)$ for $q > \qav$ and $S(q) = S_{\mathrm{th}}(q)$ for $q < \qav$.

\begin{figure}[tb]
    \includegraphics[width=\linewidth]{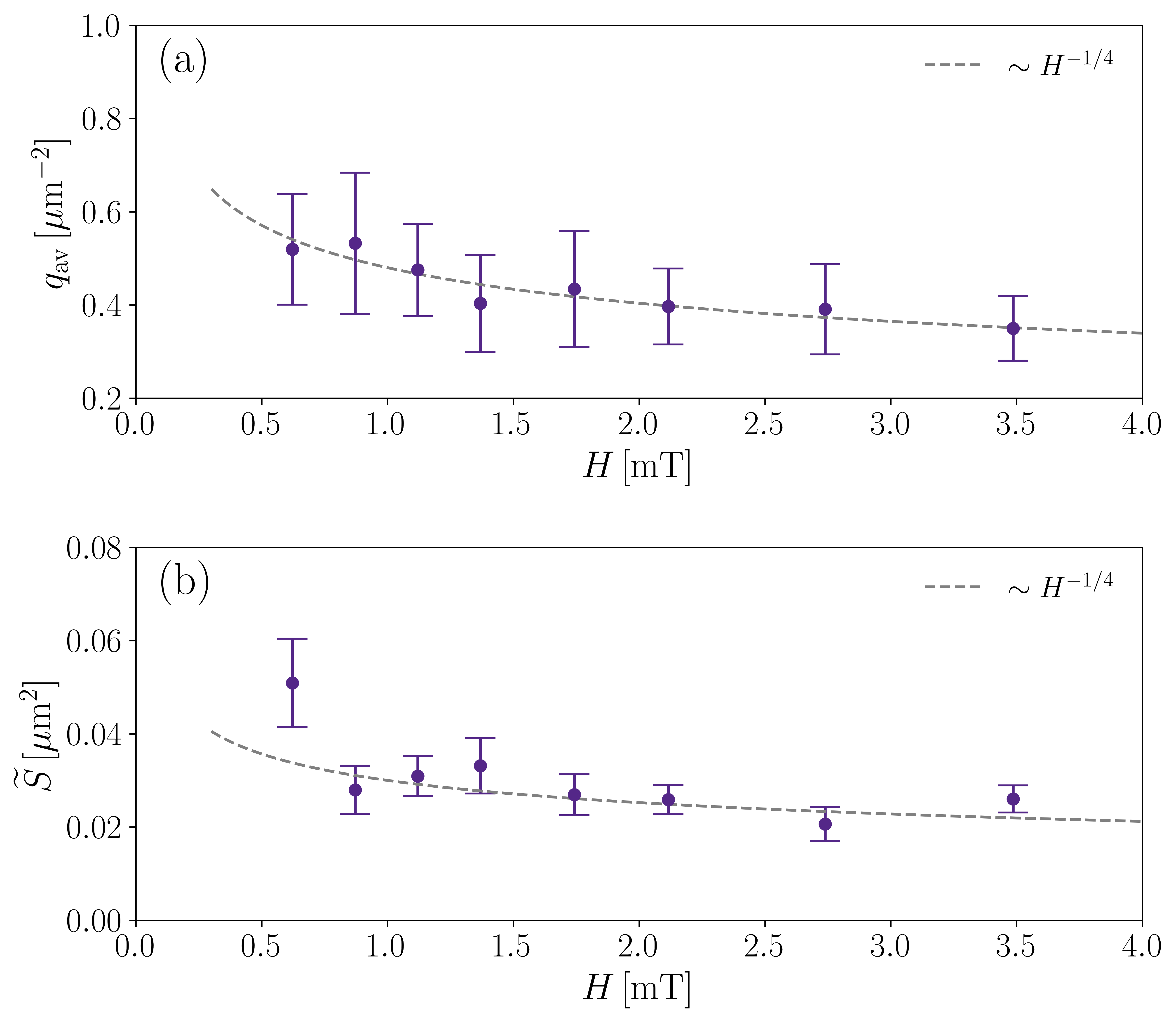}
    \caption{(a) Crossover wave vector scale $\qav$ and (b) structure factor amplitude $S_0$ as a function of the magnetic field $H$.
    The dashed lines indicate power-law behavior $\qav \sim H^{-\eta}$ and $\widetilde{S} \sim H^{-\gamma}$ (see Eq.~\eqref{eq-sigmatilde-field-gamma}) with $\eta = \gamma = 1/4$.}
    \label{fig:qavS0}
\end{figure}

Given a proposed structure factor, with given $\widetilde{S}$ and $\qav$, the associated roughness function can be written as~\cite{Albornoz2021}
\begin{equation}
\label{eq:BrSq}
    B(r) = 4 \sum_{n=1}^{(M-2)/2} S(q_n) \left[ 1 - \cos\left(q_n r \right)\right],
\end{equation}
with $M$ a large and even number. Then, considering the associated roughness function $B(r)$,  effective roughness parameters can be obtained, which depend on the arbitrarily chosen values for $\qav$ and $\widetilde{S}$. Therefore, we can search for field-dependent values of $\qav$ and $\widetilde{S}$ to recover effective roughness parameters equal to the experimental values shown in Fig.~\ref{fig:roughness-parameters}, on the same fitting range. Since the effective value of $\zeta$ does not depend on $\widetilde{S}$, we first search for the value of $\qav$ that reproduces the experimental value of $\zeta$ on the fixed range of $q$ at our disposal, at a given system size. Then, using the obtained value of $\qav$, we search for $\widetilde{S}$ to recover $B_0$. Figure~\ref{fig:qavS0} shows the obtained values of $\qav$ and $\widetilde{S}$ as a function of the applied magnetic field $H$.
Error bars for $\qav$ and $\widetilde{S}$ are obtained considering possible variations of $\zeta$ and $B_0$ (Fig.~\ref{fig:roughness-parameters}) within their own error bars.
We found that both $\qav$ and $\widetilde{S}$ slightly increase with decreasing field, which, as we show below, is closely related to the field-dependence of the roughness amplitude $B_0$ [Fig.~\ref{fig:roughness-parameters}(b)].

\begin{figure}[tb]
    \includegraphics[width=\linewidth]{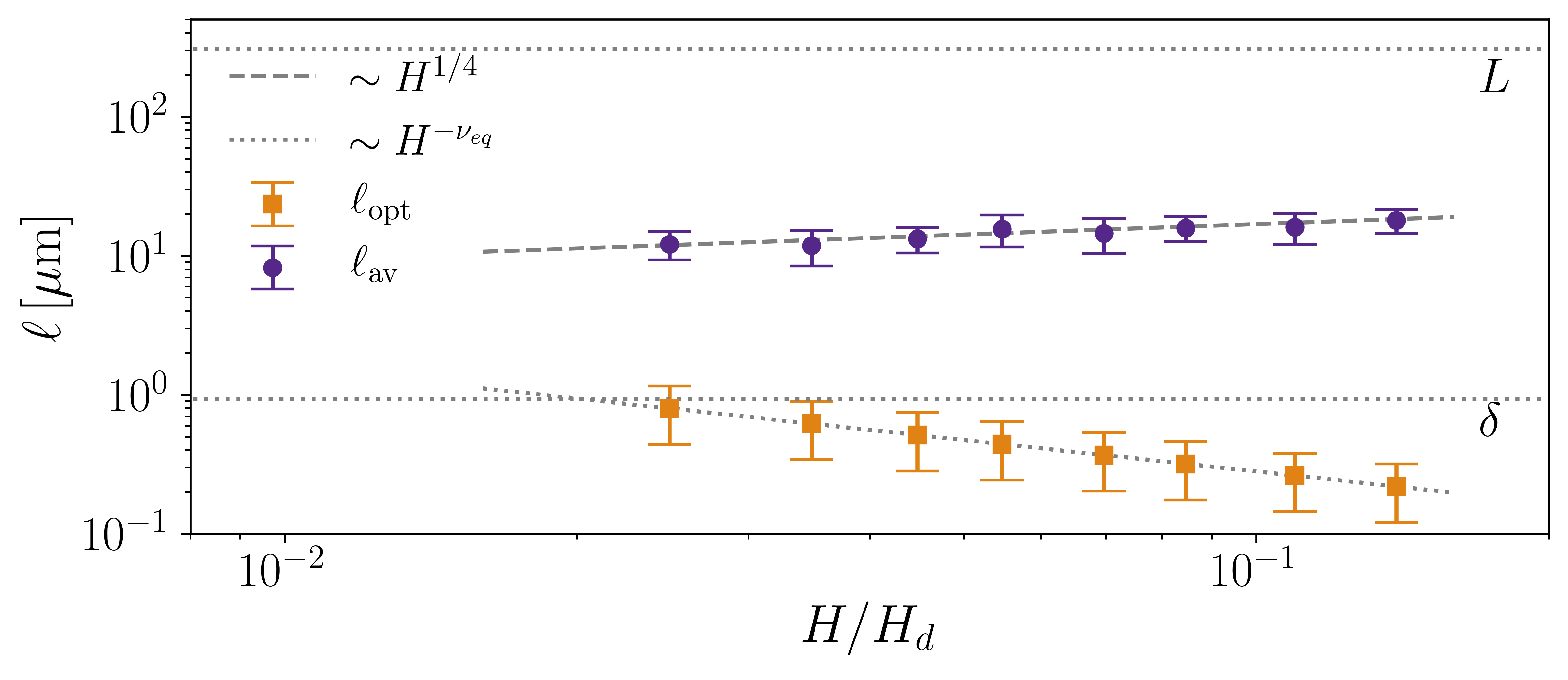}
    \caption{Length scales as functions of the scaled field $H/H_d$. $\lav = 2 \pi/\qav$ is indicated.
    The observation scales in the PMOKE experiment, $\delta$ and $L$, are displayed as dotted horizontal lines. $\lopt$ was evaluated using Eq.~\eqref{eq:lopt} with $L_c = 50\un{nm}$, $H_d = (25 \pm 15)\un{mT}$, and ${\nueq = 3/4}$ (for $\zeq=2/3$).
    }
    \label{fig:lav}
\end{figure}

Finally, Fig.~\ref{fig:lav} shows the different scales involved in the experiment. $\delta$ and $L$ correspond to the minimum and maximum of the domain wall observation scale. $\lopt$ is the optimal length scale, computed using Eq.~\eqref{eq:lopt} and estimated values for the Larkin length $L_c$ and the depinning field $H_d$, and is such that $\lopt \sim H^{-\nueq}$. The $\lav = 2 \pi/\qav$ values were obtained as the crossover length scales between depinning and thermal regimes, necessary to account for the effective roughness exponent obtained in the experiment. Phenomenologically, we found that the field dependence of $\lav$ can be approximately described by $\lav \sim H^{1/4}$ (see the discussion in Sec.~\ref{sec:disc}). As shown in Fig.~\ref{fig:lav}, $\lav$ is within the experimental range between $\delta$ and $L$ ($\lopt < \delta < \lav < L$).

\section{Scaling of the roughness amplitude}
\label{sec:scaling}

So far, we have proposed a structure factor comprising the depinning and thermal regimes and we have obtained the parameters $\qav$ and $\widetilde{S}$ that allow us to properly reproduce the experimental roughness exponent $\zeta$. In what follows, we attempt to account for the magnetic field dependence of the other roughness parameter, the roughness amplitude $B_0$.
This prompts us to connect the roughness at different scales using scaling relations. In particular we have to connect fluctuations above $\lopt$ with roughness properties below $\lopt$, characterized by $\zeq$.

Since $\lopt < \delta$, we have interpreted the observed roughness parameters as effective values obtained using the crossover structure factor between depinning roughness regime (below $\lav$) and the thermal roughness regime (above $\lav$). Below $\lopt$, the expected roughness regime is the equilibrium roughness regime, characterized by $\zeq$. Therefore, the roughness regime {at length scales $L_c < r < \lopt$, corresponding to $\qopt < q < q_c$,} can be described by the structure factor
\begin{equation}
\label{eq:Seq}
    S_{\mathrm{eq}}(q) = S_c \left( \frac{q}{q_c} \right)^{-(1+2\zeq)},
\end{equation}
with $q_c = 2 \pi/L_c$ and $S_c$ the corresponding amplitude.
Since disorder is expected to dominate for length scales above $L_c$, according to Eq.~\eqref{eq:Seq} $S_c$ is the amplitude of the structure factor precisely at $q_c$ and thus serves as a reference amplitude at the lowest scale where disorder is relevant.
Concomitantly, in the real space the roughness function, just above $L_c$ and far away from finite size effects and the crossover at $\lopt$ ($r \ll L,\,\lopt$), is expected to behave as 
\begin{equation}
    B(r)  = w^2 \left( \frac{r}{L_c}\right)^{2 \zeq},
\end{equation}
with the roughness amplitude at the Larkin length, $w^2 = J_{\mathrm{eq}} S_c$, proportional to $S_c$ (see Appendix~\ref{sec:app}). The constant $J_{\mathrm{eq}}$ depends on $L_c$ and $L$, and therefore contains information about finite size effects.

The roughness amplitude $w$ is a measure of transverse fluctuations at the Larkin length scale and is typically associated to the correlation length of the disorder~\cite{chauve_creep_long}. Considering a system at \textit{low temperatures}, in the sense that no thermal roughness contribution is present at length scales below $L_c$~\cite{agoritsas2010, agoritsas2013static}, all fluctuation scales are determined by $w$. Since according to Fig.~\ref{fig:roughness-parameters}(b) the roughness increases with decreasing field, we can assume a field dependence $w \sim H^{-\sigma}$ with a positive exponent $\sigma$ and link this behavior, using $S_c \sim w^2 \sim H^{-2\sigma}$, to the field-dependence of $\widetilde{S}$ and $B_0$ shown in Figs.~\ref{fig:roughness-parameters}(b) and~\ref{fig:qavS0}(b).

We thus have to consider three roughness regimes, corresponding to the three reference states, given by $S_{\mathrm{eq}}(q)$, $S_{\mathrm{dep}}(q)$ and $S_{\mathrm{th}}(q)$, with now two characteristic amplitudes $S_c$ and $\widetilde{S}$. These two amplitude factors can be connected by matching the structure factor at the crossover scales.
The equilibrium roughness regime is valid down to the scale  $\qopt$ and should be matched with the depinning roughness regime, which is valid for $q < \qopt$, resulting in $S_{\mathrm{eq}}(\qopt) = S_{\mathrm{dep}}(\qopt)$. From this, one finds that $\widetilde{S}$ and $S_c$ are connected through
\begin{equation}
\label{eq:S1Sc}
    \widetilde{S} = S_c \left( \frac{\qopt}{q_c} \right)^{-(1+2\zeq)} \left( \frac{\qav}{\qopt} \right)^{-(1+2\zdep)}.
\end{equation}
Therefore, {recalling that $q_c \sim 1/L_c$ is assumed to be field-independent,} for the field-dependence of $\widetilde{S}$ one has
\begin{equation}
    \widetilde{S} \sim S_c \qopt^{2(\zdep-\zeq)} \qav^{-(1+2 \zdep)} \sim H^{-\gamma}
\label{eq-sigmatilde-field-gamma}
\end{equation}
with
\begin{equation}
    \gamma = 2 \sigma-\frac{2(\zdep-\zeq)}{2-\zeq} - \eta (1+2 \zdep),
\label{eq-def-gamma-explicit}
\end{equation}
where we have used Eq.~\eqref{eq:lopt} with ${\nueq = 1/(2-\zeq)}$ for the field-dependence of $\qopt$, and $\qav \sim H^{-\eta}$ with $\eta>0$, as suggested in Figs.~\ref{fig:qavS0}(a) and~\ref{fig:lav}.

The crossover between depinning and thermal roughness regimes is given by Eq.~\eqref{eq:crossfn}, with the corresponding limiting cases given by Eqs.~\eqref{eq:Sdep} and \eqref{eq:Sth}, and helps to pinpoint the field-dependence of the roughness amplitude $B_0$. Using Eq.~\eqref{eq:crossfn} and an integral approximation to Eq.~\eqref{eq:BrSq}, we get
\begin{equation}
\label{eq:BrS0Ix}
    B(r) = 4 \widetilde{S} \frac{\qav}{q_\mathrm{min}} (\qav r)^{2 \zth} I_{\mathrm{x}}(r),
\end{equation}
where $q_\mathrm{min} = 2\pi/L$. $I_{\mathrm{x}}(r)$, on its turn, is an $r$-dependent integral such that $I_{\mathrm{x}}(r) \sim (\qav r)^{2(\zeta - \zth)}$ at small scales (see Appendix~\ref{sec:app}), with $\zeta = 0.74$ the mean experimental value shown in Fig.~\ref{fig:roughness-parameters}(a). Inserting this scaling behavior into Eq.~\eqref{eq:BrS0Ix}, the $B(r) \sim r^{2 \zeta}$ behavior is recovered with the roughness amplitude scaling as
\begin{equation}
    B_0(H) \sim \widetilde{S} \qav^{1+2 \zeta} \sim H^{-\gamma} H^{-\eta (1+2\zeta)} \sim H^{-\theta},
\end{equation}
with Eqs.~\eqref{eq-sigmatilde-field-gamma} and~\eqref{eq-def-gamma-explicit} leading to
\begin{equation}
    \theta = 2 \sigma-\frac{2(\zdep-\zeq)}{2-\zeq} - 2 \eta (\zdep - \zeta).
\end{equation}

The values of the exponents $\gamma$ and $\theta$ give the field-dependent behavior of $\widetilde{S}$ and $B_0$, respectively. In addition to the roughness exponents ($\zdep = 5/4$, $\zeq = 2/3$ and $\zeta \approx 3/4$), these values depend on $\eta$ and $\sigma$. We use for instance the value of $\eta = 1/4$, which is compatible with the results in Figs.~\ref{fig:qavS0}(a) and \ref{fig:lav}. If we assume that the roughness at the Larkin scale $w$ is independent of the field, $\sigma = 0$ results in negative values for both $\gamma$ and $\theta$, which are not consistent with our observations. Instead, assuming that $w \sim 1/H$, i.e. $\sigma = 1$, the values $\gamma = 1/4$ and $\theta = 7/8$ are obtained, which are in good agreement with the results shown in Figs.~\ref{fig:roughness-parameters}(b) and \ref{fig:qavS0}(b).

\section{Discussion}
\label{sec:disc}

We have studied the behavior of the roughness parameters of magnetic domain walls as functions of the applied magnetic field $H$. The roughness exponent $\zeta$ was interpreted as an effective value due to different roughness regimes compatible with the length scales defined by the experiment. First of all, we estimated that $\lopt \ll \delta$. We then considered only two regimes contributing to the structure factor, namely, the depinning  and the thermal (or fast-flow) regimes with exponents $\zdep$ and $\zth$ respectively. This description allowed us to account for the experimentally observed roughness exponent.
We report a finite value of $\lav$ below $H_d$ and for finite temperature, which increases with the magnetic field as $\lav \sim H^{1/4}$. This does not comply with the expected behavior of $\lav$ above $H_d$ when $T \to 0$, $\lav \sim (H-H_d)^{-\nu_\mathrm{dep}}$, although in principle we can only conjecture how $\lav$ behaves at finite temperature below the depinning field. However, although the assumption of a crossover from a depinning roughness regime to a thermal roughness regime might be incorrect, using a regime with a different roughness exponent at large length scales does not qualitatively change our main results.
Indeed, what is really needed in order to describe effective values for the experimentally obtained roughness parameters, $\zeta$ and $B_0$, is a structure factor with a crossover from the depinning roughness regime at intermediate length scales to a roughness regime with $\zeta < 1$ at large length scales.
This large scale roughness regime can be thermal, as we assumed, or it can be a different regime as, for instance, an anharmonic roughness regime with $\zeta_{\mathrm{an}} \approx 0.63$. The main results we present here do not change qualitatively if an anharmonic roughness regime is used. In principle, we could not distinguish among the two interpretations.
It is worth mentioning that we have checked that including a second crossover at $\qopt = 2 \pi/\lopt$, such that $S(q) \sim q^{-(1+2\zeq)}$ for $q \gg \qopt$, does not qualitatively change the results, giving just a slightly larger value for $\lav$ (not shown).

From our analysis, the dependence of the roughness amplitude $B_0$ on the applied field suggest that the roughness amplitude at the Larkin length $w$ effectively depends on the field. Typically, one has that $w = r_f = \max(\Delta,\xi)$, with $\Delta$ the domain wall width and $\xi$ a characteristic correlation length of the disorder~\cite{chauve_creep_long}. Originally, these quantities are thought of as constant values corresponding to the zero-field case. Our results suggest that a field dependence should be considered in order to describe the dependence of the roughness amplitude on the field.
This field dependence could be associated for instance to variations of the domain wall width related to the competition between the intrinsic disorder and the external field, an ingredient currently missing from theoretical modeling.
For instance, numerical simulations of such one-dimensional interface usually rely on the so-called quenched Edwards-Wilkinson dynamics~\cite{Ferrero2013}, an effective description which depends on a few effective parameters: an elastic energy per unit of length $c$, the amplitude of the disordered potential $D$ and its correlation length $\xi$, as well as the interface width $\Delta$. Those parameters are typically assumed to be constants, but a proper microscopic justification to this \textit{ad hoc} assumption is still lacking. 
Nothing prevents us from assuming that they might in fact display a temperature or field dependence. In that respect, recent studies on how to relate this effective one-dimensional description to a two-dimensional Ginzburg-Landau description~\cite{caballero2020bulk} and further to the micromagnetic Landau-Lifshitz-Gilbert model~\cite{Guruciaga2021} pave the way to exploring such additional dependence on a firmer basis.

Finally, it is worth to mention that in our current analysis we have considered that the Larkin length $L_c$ does not depend on the field. However, once we question the possible field-dependence of $w$, the roughness amplitude at the scale $L_c$, one might wonder whether $L_c$ could also depend on the field. If both $w$ and $L_c$ depend on $H$, the behavior $B_0 \sim H^{-\theta}$ is recovered considering that $(w^2 L_c) \sim H^{-2 \sigma}$. A much detailed study of the field dependence of the disorder-dependent parameters is thus needed.

\section{Conclusion}
\label{sec:conc}

In summary, we reported measured roughness parameters and their field dependence in the creep regime for a Pt/Co/Pt magnetic thin film. We found agreement between mean values obtained from individual profiles and values from the average roughness function. This is the first experimental report of roughness parameters obtained using an average roughness function $\langle B(r) \rangle$; previous results (see data in Ref.~\cite{Albornoz2021}) were obtained fitting the roughness parameters to individual $B(r)$.

The roughness amplitude $B_0$ increases with decreasing field, pointing to rougher domain walls for lower field values, while the roughness exponent $\zeta$ is approximately field-independent. This confirms that the roughness variation observed by naked eye when changing the applied magnetic field is fully due to the roughness amplitude (as previously observed~\cite{metaxas_depinning_thermal_rounding}). 
In addition, we found that a finite $\lav$ signaling the crossover between small scale behavior with $\zeta = \zdep$ and large scale behavior with $\zeta = \zth$ can account for experimentally obtained roughness exponents close to $\zeta = 0.74$. These effective exponents would then also depend --to a certain extent-- on the typical scales used in the experiments, $\delta$ and $L$.

In order to account for the experimentally observed roughness exponent, a crossover between a depinning roughness regime and a roughness regime with $\zeta < 1$ at larger scales is needed. In addition, a dependence with the field of the roughness amplitude at the Larkin length is required to describe the field-dependence of the roughness amplitude $B_0$, a commonly observed behavior.

\begin{acknowledgments}
    We thank L.~J.~Albornoz for useful discussions. This work was supported by Agencia Nacional de Promoci\'on Cient\'\i fica y Tecnol\'ogica (PICT 2016-0069, PICT 2017-0906 and PICT 2019-02873) and Universidad Nacional de Cuyo (grants 06/C561 and M083). E.A. acknowledges support from the Swiss National Science Foundation (SNSF) by the SNSF Ambizione Grant PZ00P2{\_}173962.
\end{acknowledgments}

\appendix
\section{Relation between $B(r)$ and $S(q)$ amplitudes}
\label{sec:app}

In order to relate the amplitudes accompanying structure factor and roughness function to important theoretical quantities at small scales, we first have to compute the roughness function at the Larkin scale.
First, a relationship between $B(r)$ and $S(q)$ amplitudes should be established. We shall rely on the continuous approximation to Eq.~\eqref{eq:BrSq}, obtained by transforming the sum into an integral using $\Delta n \to \Delta q/(2 \pi/L)$:
\begin{equation}
\label{eq:intBrSq}
    B(r) = \frac{4}{q_\mathrm{min}} \int_{q_\mathrm{min}}^{q_\mathrm{max}} dq [1-\cos(qr)] S(q),
\end{equation}
with the integral running between $q_\mathrm{min} = 2\pi/L$ and $q_\mathrm{max} = \pi /\delta$. Notice that both $S(q)$ and $B(r)$ have dimensions of length squared.
Notice too that the $1/q_\mathrm{min}$ appears due to the definition of the structure factor we are using and guaranties proper dimensions for $B(r)$ and $S(q)$.

\begin{figure}[tb]
    \includegraphics[width=\linewidth]{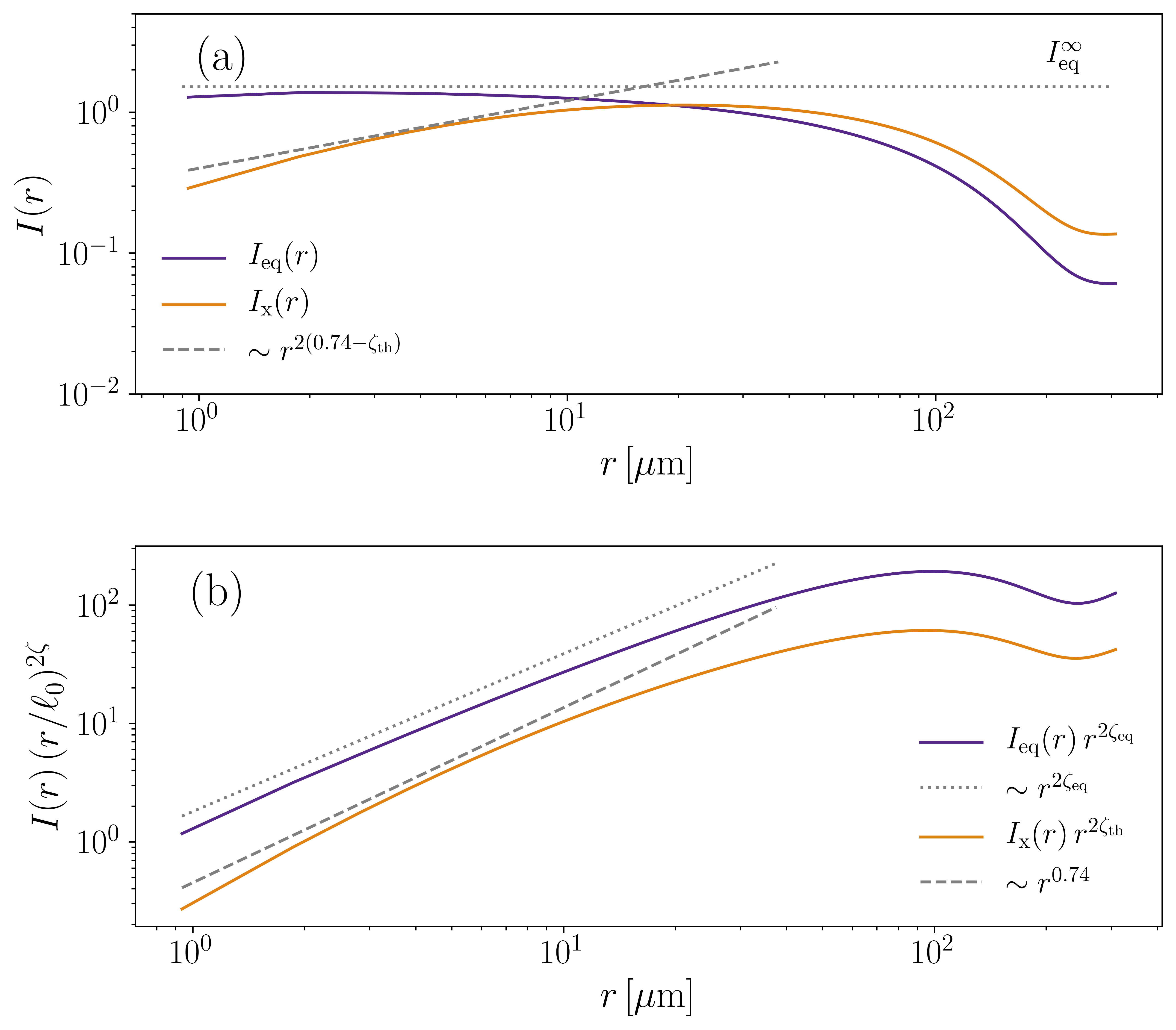}
    \caption{(a) Dependence with $r$ of the integrals $I_{\mathrm{eq}}(r)$ and $I_{\mathrm{x}}(r)$ given by Eqs.~\eqref{eq:Ieq} and~\eqref{eq:Ix}, respectively.
    (b) By multiplying these integrals by $r^{2 \zeta}$, with the corresponding $\zeta$ value, we can appreciate how the integrals correct the $B(r)$ behavior.
    To numerically compute $I_{\mathrm{x}}(r)$ the value $\qav = 0.4\un{\mu m^{-2}}$ was used.}
    \label{fig:Ir}
\end{figure}

The connection between the roughness and structure factor amplitudes can then be illustrated by first considering a finite system with a very large $\lopt$ such that $L_c < \delta < L < \lopt$, whose position fluctuations are described by equilibrium roughness only.
Using Eqs.~\eqref{eq:Seq} and \eqref{eq:intBrSq} and an adimensional parameter $v = q r$, we can write
\begin{equation}
B(r) = 4 S_c\frac{q_c}{q_\mathrm{min}} I_{\mathrm{eq}}(r) \, (q_c r)^{2 \zeq} \end{equation}
with $q_c = 2 \pi/L_c$ and
\begin{equation}
\label{eq:Ieq}
    I_{\mathrm{eq}}(r) = \int_{v_\mathrm{min}(r)}^{v_\mathrm{max}(r)} dv (1-\cos v) v^{-(1+2 \zeq)},
\end{equation}
where $v_\mathrm{min}(r) = q_\mathrm{min} r$, $v_\mathrm{max}(r) = q_\mathrm{max} r$. Notice that the integral $I_{\mathrm{eq}}(r)$ contains a $r$-dependence through its limits. This results in a small finite size correction to the roughness exponent. Figure~\ref{fig:Ir}(a) shows a numerical calculation of the integral $I_{\mathrm{eq}}(r)$. In addition, Fig.~\ref{fig:Ir}(b) presents the product $I_{\mathrm{eq}}(r) r^{2 \zeq}$, which determines the $r$-dependence of $B(r)$. Since $I_{\mathrm{eq}}(r)$ displays an almost constant behavior at small scales, it is expected that $B(r) \sim r^{2 \zeq}$.
For an infinite system the integral tends to the constant value $I_{\mathrm{eq}}^\infty = \Gamma(-4/3)/2 \approx 1.52$, shown as a dotted horizontal line in Fig.~\ref{fig:Ir}(a), and the roughness exponent is exactly $\zeq$. Then, in this case, the roughness function can be expressed as
\begin{equation}
    B(r)  = w^2 \left( \frac{r}{L_c}\right)^{2 \zeq},
\end{equation}
with
\begin{equation}
    w^2 = 4 S_c (2 \pi)^{2 \zeq} I_{\mathrm{eq}}^\infty \frac{q_c}{q_\mathrm{min}} = J_{\mathrm{eq}} S_c.
\end{equation}
Therefore, at the scale $L_c$, the roughness amplitude $w^2$ is proportional to $S_c$ and if $w \sim H^{-\sigma}$ then $S_c \sim H^{-2 \sigma}$. The constant $J_{\mathrm{eq}}$ depends on $L_c$ and $L$, and therefore contains information about finite size effects. Using $L_c = 50\un{nm}$ and $L=309\un{\mu m}$, we obtain $J_{\mathrm{eq}} \sim 2 \times 10^5$ as an order of magnitude.

Lets us now consider the crossover between thermal and depinning roughness regimes given by Eq.~\eqref{eq:crossfn}, with the corresponding limiting cases given by Eqs.~\eqref{eq:Sdep} and~\eqref{eq:Sth}. Using Eqs.~\eqref{eq:crossfn} and~\eqref{eq:intBrSq}, we get
\begin{equation}
    B(r) = 4 \widetilde{S} \frac{\qav}{q_\mathrm{min}} (\qav r)^{2 \zth} I_{\mathrm{x}}(r),
\end{equation}
where
\begin{equation}
\label{eq:Ix}
    I_{\mathrm{x}}(r) = \int_{v_\mathrm{min}(r)}^{v_\mathrm{max}(r)} dv (1-\cos v) G(v,r)
\end{equation}
with
\begin{equation}
    G(v,r) = \frac{v^{-(1+2 \zth)}}{1+(\qav r)^{-2(\zdep-\zth)} v^{2(\zdep-\zth)}}.
\end{equation}
The integral $I_{\mathrm{x}}(r)$ depends on $r$ through its limits, resulting in finite size effects, and through the factor $\qav r$ in the function $G(v,r)$ inside it. These dependencies and the $r^{2 \zth}$ factor give rise to the effective value of the roughness exponent. A numerical calculation of $I_{\mathrm{x}}(r)$ is shown in Fig.~\ref{fig:Ir}(a). At small scales it presents a $r$-dependence which, when multiplied by $r^{2 \zth}$, should result in $B(r) \sim r^{2 \zeta}$, as shown in Fig.~\ref{fig:Ir}(b). Accordingly, as shown in Fig.~\ref{fig:Ir}(a), $I_{\mathrm{x}}(r)$ behaves approximately as $r^{2(\zeta - \zth)}$ at small scales.

\bibliography{biblio}

\end{document}